\documentstyle[12pt]{article}
\newcommand{\NP}{Nucl. Phys. }

\newcommand{\PR}{Phys. Rev. }
\newcommand{\PRL}{Phys. Rev. Lett. }
\newcommand{\PL}{Phys. Lett. }

\newcommand{\calL}{{\cal L}}

\newcommand{\lam}{\lambda}

\newcommand{\veca}{{\bf a}}

\newcommand{\vecB}{{\bf B}}

\begin{document}
\baselineskip=16pt

\pagenumbering{arabic}

\vspace{1.0cm}

\begin{center}
{\Large\sf Impact of spin-zero particle-photon interactions on
light polarization in external magnetic fields}
\\[10pt]
\vspace{.5 cm}

{Yi Liao\footnote{liaoy@nankai.edu.cn}}

{\small Department of Physics, Nankai University, Tianjin 300071,
China} %

\vspace{2.0ex}

{\bf Abstract}

\end{center}

If the recent PVLAS results on polarization changes of a linearly
polarized laser beam passing through a magnetic field are
interpreted by an axion-like particle, it is almost certain that
it is not a standard QCD axion. Considering this, we study the
general effective interactions of photons with spin-zero particles
without restricting the latter to be a pseudo-scalar or a scalar,
i.e., a parity eigenstate. At the lowest order in effective field
theory, there are two dimension-5 interactions, each of which has
previously been treated separately for a pseudo-scalar or a scalar
particle. By following the evolution in an external magnetic field
of the system of spin-zero particles and photons, we compute the
changes in light polarization and the transition probability for
two experimental set-ups: one-way propagation and round-trip
propagation. While the first may be relevant for astrophysical
sources of spin-zero particles, the second applies to laboratory
optical experiments like PVLAS. In the one-way propagation,
interesting phenomena can occur for special configurations of
polarization where, for instance, transition occurs but light
polarization does not change. For the round-trip propagation,
however, the standard results of polarization changes for a
pseudoscalar or a scalar are only modified by a factor that
depends on the relative strength of the two interactions.

\begin{flushleft}
% corresponding to: (PACS 2006 version)
% Axions and other Nambu-Goldstone bosons; Experimental tests; Birefringence,
% polarization

\end{flushleft}

\newpage

The neutral photon can interact with itself due to quantum
effects. In QED this is summarized in the celebrated
Euler-Heisenberg Lagrangian for weak external fields and low
frequency photons \cite{heisenberg}. Among physical effects of
photon field self-interactions is the nonlinear optical phenomenon
of a light beam in a magnetized vacuum \cite{adler70,iacopini79}.
This latter effect has long been searched for in laboratory
experiments without success \cite{BFRT} until recently by PLVAS,
in which the rotation of the plane of polarization (dichroism)
\cite{PVLAS05} and ellipticity (birefringence) \cite{PVLAS06} of a
linearly polarized laser beam are observed after it traverses a
transverse strong magnetic field. It is amusing that the observed
rotation and ellipticity exceed the QED expectation by many orders
of magnitude \cite{adler70,adler06,biswas06}. Without restricting
to QED but allowing all possible photon field self-interactions,
it has been shown that it is still not possible to accommodate the
PVLAS results of rotation and ellipticity at the first non-trivial
order in the low energy theory of photons \cite{hu07}. This would
mean that some ultra-light particles with a mass of order the
laser frequency or lower and interacting with photon fields have
to be invoked. The real and virtual production of these particles
by laser in the external field induces the optical changes of the
initial beam \cite{sikivie83,maiani86,raffelt88}. If the PVLAS
results are confirmed, it would signal some new physics containing
ultra-light particles.

The theoretically best motivated candidate for such a particle
seems to be the axion, a pseudoscalar particle resulting from
spontaneous breakdown of the Peccei-Quinn symmetry. In addition to
offering a natural resolution to the strong CP problem, the axion
also serves as an attractive candidate for dark matter. However,
its mass and coupling to photons as determined by PVLAS are
strongly excluded by helioscope experiments \cite{cast05,cast07}
and astrophysical observations \cite{raffelt06}. There are many
discussions in the literature on this potential conflict, most
appealing to hitherto unknown effects in the stellar environment
\cite{conflict}. An alternative to the axion or axion-like
particles is the milli-charged particles with mass of order
$0.1~{\rm eV}$ and an electric charge of order $10^{-6}~e$,
interacting with photons as in QED \cite{milli}. These particles
were shown some years ago to appear in para-photon models through
kinetic mixing \cite{holdom86}, and have recently been argued to
appear naturally in string-based models \cite{abel06}. They have
also been utilized to circumvent the above mentioned conflict
\cite{masso06}, and may have detectable effects in accelerator
cavities \cite{gies06} and reactor neutrino experiments
\cite{Gninenko06}. Whether any of these suggestions is relevant
will be examined in the new generation of experiments; in
particular, axion-like particles can be best scrutinized in photon
regeneration experiments \cite{sikivie83,BFRT,regen_e,regen_t},
one of which will start to operate soon by the ALPS group at DESY
\cite{alps}.

It is not the purpose of this work to add another proposal to
relax the tension between the PVLAS and other experiments.
Instead, our work is based on the observation \cite{raffelt06}
that, if the PVLAS results are interpreted in terms of an
axion-like particle, it cannot be a QCD axion because the implied
relation between mass and coupling is far away from the parameter
region favored by PVLAS. In this circumstance there is no
theoretical prejudice for a pseudoscalar or a scalar to explain
the PVLAS results. In the effective theory at the low energy scale
($\sim 1~\rm{eV}$), the spin-zero particle probed by PVLAS could
well be one of no definite parity, i.e., a mixture of pseudoscalar
and scalar. The portion of mixing depends on the details of some
underlying fundamental theory that produces the ultra-light
particle. Such particles could appear in extensions of the
standard model as studied, for instance, in Ref.\cite{hill88}.
Then we should include in our effective theory all possible
interactions of the particle with the photon field that have the
same dimension. Note that there is no direct threat to experiments
on parity and CP violation at much higher energy scales where our
effective theory does not generally apply.

The effective theory at the lowest non-trivial order is defined by
\begin{eqnarray}
\calL&=&-\frac{1}{4}F^{\mu\nu}F_{\mu\nu}
+\frac{1}{2}(\partial^{\mu}\varphi)^2-\frac{1}{2}m^2\varphi^2
+\frac{1}{4}\lam_+\varphi F^{\mu\nu}F_{\mu\nu}
+\frac{1}{4}\lam_-\varphi \tilde{F}^{\mu\nu}F_{\mu\nu},
\end{eqnarray}
where $\varphi$ is a spin-zero field, $F_{\mu\nu}$ the
electromagnetic field tensor with dual
$\tilde{F}^{\mu\nu}=\frac{1}{2}\epsilon^{\mu\nu\alpha\beta}F_{\alpha\beta}$
($\epsilon^{0123}=+1$), and $\lam_{\pm}$ are two coupling
constants. The separate cases with either $\lam_+=0$ ($\varphi$
being a pseudoscalar) or $\lam_-=0$ (a scalar) have been studied
in the literature. The equations of motion (EoM's) are
\begin{eqnarray}
(\partial^2+m^2)\varphi&=&
\frac{1}{4}\lam_+F^{\mu\nu}F_{\mu\nu}
+\frac{1}{4}\lam_-\tilde{F}^{\mu\nu}F_{\mu\nu}\nonumber\\
\partial_{\mu}F^{\mu\nu}&=&\partial_{\mu}\left(\lam_+\varphi
F^{\mu\nu}+\lam_-\varphi\tilde{F}^{\mu\nu}\right)
\end{eqnarray}
Since the background classical field will be much larger in
strength than the quantum ones, the leading effects to quantum
fields can be approximated by linearizing the EoM's with respect
to them:
\begin{eqnarray}
(\partial^2+m^2)\varphi&=&\frac{1}{2}\lam_+f_{\mu\nu}F^{\mu\nu}
+\frac{1}{2}\lam_-f_{\mu\nu}\tilde{F}^{\mu\nu}\nonumber\\
\partial_{\mu}f^{\mu\nu}&=&
\left(\lam_+F^{\mu\nu}+\lam_-\tilde{F}^{\mu\nu}\right)\partial_{\mu}\varphi
\end{eqnarray}
where $F~(f)$ stands for the background (laser light) field
tensor, and Maxwell equations have been used to simplify the
second equation. Denoting
$f_{\mu\nu}=\partial_{\mu}a_{\nu}-\partial_{\nu}a_{\mu}$ and
working in the gauge $a^0=\nabla\cdot{\bf a}=0$, the EoM's in an
external magnetic field ${\bf B}$ simplify to
\begin{eqnarray}
(\partial_t^2-\nabla^2+m^2)\varphi&=&
\lam_+(\nabla\times\veca)\cdot\vecB+\lam_-(\partial_t\veca)\cdot\vecB
\nonumber\\
(\partial_t^2-\nabla^2)\veca&=&
\lam_+(\nabla\varphi)\times\vecB-\lam_-(\partial_t\varphi)\vecB
\end{eqnarray}

We follow Ref.\cite{raffelt88} to derive the evolution equation
for a plane-wave light beam propagating in the $z$ direction and
perpendicularly to ${\bf B}$. We can choose $a^3=0$, and
$a^1,a^2,\varphi$ depend only on $(t,z)$. Since for a temporally
constant ${\bf B}$, the energies of the laser photon and the
$\varphi$ particle are equal, we can remove $t$-dependence by
substitutions ${\bf a}\to{\bf a}e^{-i\omega t},~\varphi\to\varphi
e^{-i\omega t}$, where $\omega$ is the angular frequency of the
laser light. For propagation in the $+z$ direction, we can
substitute on the left-hand side of equations,
$(\omega^2+\partial_z^2)=(\omega-i\partial_z)(\omega+i\partial_z)
\to 2\omega(\omega+i\partial_z)$, and on the right,
$\partial_z\to+i\omega$. The resultant errors are higher order in
small parameters not to be considered here. The EoM's in the
matrix form are
\begin{equation}
\left(1+i\omega^{-1}\partial_z+\Omega\right)
\left(\begin{array}{c}a^1\\a^2\\\varphi\end{array}\right)=0,~~
\Omega=\left(\begin{array}{ccc}
0&0&+i\delta_-\\0&0&+i\delta_+\\
-i\delta_-&-i\delta_+&-\delta_0\end{array}\right), \label{linear}
\end{equation}
where
\begin{equation}
\delta_-=\frac{1}{2\omega}(\lam_-B_1-\lam_+B_2),~
\delta_+=\frac{1}{2\omega}(\lam_-B_2+\lam_+B_1),~
\delta_0=\frac{m^2}{2\omega^2}
\end{equation}
Without loss of generality, we choose $\vecB=|\vecB|\hat{x}$, then
$\delta_{\pm}=(2\omega)^{-1}\lam_{\pm}|\vecB|$. In this coordinate
system, $a^{1,2}$ are respectively the components parallel and
perpendicular to $\vecB$, thus denoted below as
$a_{\parallel,\perp}$.

To find the eigenmodes of propagation, we diagonalize the matrix
$\Omega$ in steps, with the result
\begin{eqnarray}
\Omega_{\rm diag}&=&U^{-1}\Omega U=\delta_0~{\rm diag}
\left(0,\epsilon^2,-(1+\epsilon^2)\right)
+\delta_0O(\epsilon^4),\\
U&=&ER_a(\theta_{\lam})R_{a\varphi}(\epsilon)=E\left(\begin{array}{rrr}
c_{\lam}&s_{\lam}c_{\epsilon}&-s_{\lam}s_{\epsilon}\\
-s_{\lam}&c_{\lam}c_{\epsilon}&-c_{\lam}s_{\epsilon}\\
0&s_{\epsilon}&c_{\epsilon}
\end{array}\right)
\end{eqnarray}
where $E={\rm diag}(i,i,1)$ removes the $\pm i$ factors in
$\Omega$, $R_a(\theta_{\lam})$ rotates the two components
$\delta_{\pm}$ into one,
\begin{equation}
\delta=\sqrt{\delta^2_++\delta^2_-}=\frac{|{\bf
B}|}{2\omega}\sqrt{\lam^2_++\lam^2_-}
\end{equation}
which in turn mixes with $\varphi$. This last mixing is then
diagonalized by $R_{a\varphi}(\epsilon)$. The following short-cuts
\begin{eqnarray}
s_{\lam}=\sin\theta_{\lam}=\frac{\lam_-}{\sqrt{\lam_+^2+\lam_-^2}},
~c_{\lam}=\cos\theta_{\lam}=\frac{\lam_+}{\sqrt{\lam_+^2+\lam_-^2}}\\
s_{\epsilon}=\sin\epsilon,~c_{\epsilon}=\cos\epsilon,~\tan
2\epsilon=\frac{2\delta}{\delta_0}= \frac{2\omega|{\bf
B}|}{m^2}\sqrt{\lam_+^2+\lam_-^2}
\end{eqnarray}
have been used. For PVLAS one has $\omega=\frac{2\pi}{\lam}\approx
1.17~{\rm eV}$, $|\vecB|=5~{\rm Tesla}\approx 976.68~{\rm eV}^2$.
PVLAS found that one would want $\lam_{\pm}\sim (4\times 10^5~{\rm
GeV})^{-1},~m\sim 10^{-3}~{\rm eV}$ to explain the results. Thus
the above mixing is very small, $\epsilon\sim 3\times 10^{-6}$.

A state that enters the ${\bf B}$ field at $z_0$,
$\Psi(z_0)=(a_{\parallel}(z_0),a_{\perp}(z_0),\varphi(z_0))^T$
will evolve into the state,
$\Psi(z)=(a_{\parallel}(z),a_{\perp}(z),\varphi(z))^T$, after
traversing a distance $z-z_0=L\ge 0$ in the $+z$ direction:
\begin{equation}
\Psi(z_0+L)=e^{i\omega L}V_+(L)\Psi(z_0),~ V_+(L)=Ue^{i\omega
L\Omega_{\rm diag}}U^{-1}
\end{equation}
Expansion in $\epsilon$ yields
\begin{equation}
V_+(L)=V_0(L)+\epsilon V_1(L)+\epsilon^2 V_2(L)+O(\epsilon^3)
\end{equation}
where
\begin{eqnarray}
V_0(L)&=&{\rm diag}(1,1,e^{-i\zeta}),~\zeta=\delta_0\omega L\\
V_1(L)&=&(1-e^{-i\zeta})\left(\begin{array}{ccc}
&&is_{\lam}\\&&ic_{\lam}\\-is_{\lam}&-ic_{\lam}&0
\end{array}\right)\\
V_2(L)&=&(-1+i\zeta+e^{-i\zeta}) \left(\begin{array}{ccc}
s_{\lam}^2&c_{\lam}s_{\lam}&\\
c_{\lam}s_{\lam}&c_{\lam}^2&\\
&&0\end{array}\right)\nonumber\\
&+&(1-e^{-i\zeta}(1+i\zeta)){\rm diag}(0,0,1)
\end{eqnarray}
For $\lam_+=0$, we have $c_{\lam}=0,~s_{\lam}={\rm sign}(\lam_-)$
and $s_{\lam}\epsilon=\omega|\vecB|\lam_-m^{-2}$. This is the case
elaborated upon in Ref.\cite{raffelt88}. The difference to that
reference is that the $\pm i$ factors multiplying
$s_{\lam},~c_{\lam}$ in $V_1(z)$ (which amount to a re-phasing of
the photon state) were dropped there and that their $\zeta$ seems
to be $-\zeta$ for propagation in the $+z$ direction.

For propagation in the $-z$ direction, the linearized EoM's can be
obtained from eqn. (\ref{linear}) by $\partial_z\to -\partial_z$
and $\lam_+\to -\lam_+$. The latter replacement amounts to
$c_{\lam}\to -c_{\lam}$ in the matrices $U,~V$ for our choice
${\bf B}=|{\bf B}|\hat{x}$. The former has no effect as long as
the phase retardation is measured by the distance $L\ge 0$. Thus,
a state, $\Psi(z_0)$, entering the ${\bf B}$ field at $z_0$ will
evolve into the following one, after traversing a distance
$z_0-z=L$ in the $-z$ direction:
\begin{equation}
\Psi(z_0-L)=e^{i\omega L}V_-(L)\Psi(z_0),~
V_-(L)=V_+(L)|_{c_{\lam}\to -c_{\lam}}
\end{equation}

Now we consider the optical effects of the evolution equation.
Suppose a beam of light propagates in the $+z$ direction and
enters a transverse ${\bf B}={\bf B}\hat{x}$ field. If the initial
beam is linearly polarized at an angle $\theta$ measured
counter-clockwise in the $(xy)$ plane with respect to the field,
it will evolve into the following state after traversing a
distance $L$ in the field:
\begin{equation}
\left(\begin{array}{l}a_{\parallel}\\a_{\perp}\\\varphi
\end{array}\right)(L)=e^{i\omega L}V_+(L)
\left(\begin{array}{l}c_{\theta}\\s_{\theta}\\0
\end{array}\right)\equiv e^{i\omega L}\left(
\begin{array}{c}
\eta\cos(\theta+\Delta\theta)e^{i\phi_{\parallel}}\\
\eta\sin(\theta+\Delta\theta)e^{i\phi_{\perp}}\\
\rho e^{i\sigma}
\end{array}\right)
\end{equation}
where, to $O(\epsilon^2)$,
\begin{eqnarray}
\eta\cos(\theta+\Delta\theta)&=&\cos\theta-
\epsilon^2(1-\cos\zeta)\sin(\theta_{\lam}+\theta)\sin\theta_{\lam}
\nonumber\\
\eta\sin(\theta+\Delta\theta)&=&\sin\theta-
\epsilon^2(1-\cos\zeta)\sin(\theta_{\lam}+\theta)\cos\theta_{\lam}
\nonumber\\
\phi_{\parallel}&=&
\epsilon^2\sin(\theta_{\lam}+\theta)\frac{\sin\theta_{\lam}}
{\cos\theta}(\zeta-\sin\zeta)\\
\phi_{\perp}&=&\epsilon^2\sin(\theta_{\lam}+\theta)\frac{\cos\theta_{\lam}}{\sin\theta}
(\zeta-\sin\zeta)\nonumber\\
\rho e^{i\sigma}&=&
2\epsilon\sin(\theta_{\lam}+\theta)\sin\frac{\zeta}{2}
\exp\left(-\frac{i}{2}\zeta\right)\nonumber
\end{eqnarray}
Conservation of probability requires $\eta^2+\rho^2=1$ as can be
checked explicitly.

The polarization of the initial beam has been rotated at
$O(\epsilon^2)$ by,
\begin{equation}
\Delta\theta=-\epsilon^2\sin^2\frac{\zeta}{2}\sin
2(\theta_{\lam}+\theta) \label{rotation}
\end{equation}
The relative phase shift of the parallel and perpendicular
components of the light beam results in the ellipticity measured
by $\tan\chi$ where $\chi\in[-\frac{\pi}{4},\frac{\pi}{4}]$ is
determined by $\sin 2\chi=\sin
2(\theta+\Delta\theta)\sin(\phi_{\perp}-\phi_{\parallel})$. Thus,
at $O(\epsilon^2)$, we have
\begin{equation}
\tan\chi\approx\chi\approx\frac{1}{2}\epsilon^2(\zeta-\sin\zeta)\sin
2(\theta_{\lam}+\theta) \label{ellipticity}
\end{equation}
Although the relative phase shift has to be treated carefully at
the angles with $\sin 2\theta=0$, the above ellipticity is always
well defined. The production probability of $\varphi$ is
\begin{equation}
P[(\gamma{\rm ~at~}\theta)\to\varphi]=|\rho|^2=4\epsilon^2
\sin^2(\theta_{\lam}+\theta)\sin^2\frac{\zeta}{2}
\end{equation}
The familiar results for a pure pseudoscalar or scalar are
recovered by putting $c_{\lam}=0,~|s_{\lam}|=1$ or
$s_{\lam}=0,~|c_{\lam}|=1$ respectively.

In a photon regeneration or shinning-light-through-walls
experiment like ALPS, $\varphi$ particles would be produced by
laser in the production zone, then penetrate a wall that blocks
the laser, and enter into the detection zone. This beam of
$\varphi$ particles will evolve into the state
\begin{equation}
e^{i\omega L}\left(i\epsilon s_{\lam}(1-e^{-i\zeta}), i\epsilon
c_{\lam}(1-e^{-i\zeta}),
e^{-i\zeta}+\epsilon^2[1-e^{-i\zeta}(1+i\zeta)]\right)^T
\end{equation}
The probability to produce a photon is thus
\begin{equation}
P[\varphi\to{\rm photon}]=4\epsilon^2\sin^2\frac{\zeta}{2}
\end{equation}
The produced photon is linearly polarized at the angle $\theta$ to
the magnetic field, determined by $\tan\theta\tan\theta_{\lam}=1$.
It would thus be possible to extract the parity property of
$\varphi$ if the photon's polarization could be measured.

A few comments concerning the above results are in order. When
$\sin(\theta_{\lam}+\theta)=0$, i.e., $\tan\theta=-\lam_-/\lam_+$,
there are no optical effects and a photon with this particular
polarization relative to $\vecB$ cannot be converted into a
$\varphi$, although an existing $\varphi$ can still be converted
into a photon with an orthogonal polarization. When
$\cos(\theta_{\lam}+\theta)=0$, i.e., $\tan\theta=\lam_+/\lam_-$,
there are no optical effects due to equal attenuation and phase
retardation of the two orthogonal polarizations, but the photon to
$\varphi$ transition probability reaches its maximum which is
equal to that of the inverse transition. Finally, for
$\zeta=\frac{m^2L}{2\omega}\ll 1$, the two optical quantities
induced by a spin-zero particle are simply related by
\begin{equation}
\frac{\Delta\theta}{\chi}=-\frac{3}{\zeta}=-\frac{6\omega}{m^2L}
\end{equation}

While the above results may be relevant for astrophysical sources
of $\varphi$ particles, they are insufficient for laboratory
optical experiments where the laser light is reflected back and
forth in a magnetic field for a larger gain of the signal. Due to
parity violating interactions, the changes of polarization in
opposite directions are not equal, and can cancel partially for a
round trip. The detailed information on the evolution of a state
$\Psi(0)$ can be obtained as follows for a round trip in the
$\vecB$ field. Suppose it propagates first for a distance $L$ in
the $+z$ direction and perpendicularly to $\vecB$, gets reflected
by a high-reflectivity mirror, then propagates in the $-z$
direction for the same distance and is reflected back to its
starting point. The state becomes
\begin{equation}
\Psi_{\rm round}=(Re^{i\omega L}V_-)(Re^{i\omega L}V_+)\Psi(0)
\end{equation}
where $R={\rm diag}(1,1,0)$ expresses the fact that the produced
$\varphi$ particles are not reflected but penetrate the mirror to
disappear. The evolution is thus not coherent at the mirrors.
Irrelevant global phases at reflection have been neglected in the
above.

In all optical experiments the light is reflected many times.
Since the signal is extremely small for one passage in the
magnetic field, the final state can be well approximated by that
after $N$ times of round trips:
\begin{equation}
\Psi_{N,{\rm round}}=e^{i2\omega NL}(RV_-RV_+)^N\Psi(0)
\end{equation}
where $N$ is roughly half the total number of passage of light in
the field. Using
\begin{equation}
(RV_- RV_+)^N=\left(\begin{array}{ccc} C^N&0&C^{N-1}X\\
0&S^N&S^{N-1}Y\\0&0&0\end{array}\right)
\end{equation}
where, to $O(\epsilon^2)$,
\begin{equation}
\begin{array}{rcl}
C&=&1+2\epsilon^2s_{\lam}^2(-1+i\zeta+e^{-i\zeta})\\
&\approx&[1-4\epsilon^2s_{\lam}^2\sin^2\frac{\zeta}{2}]
e^{i2s_{\lam}^2\epsilon^2(\zeta-\sin\zeta)}\\
S&=&1+2\epsilon^2c_{\lam}^2(-1+i\zeta+e^{-i\zeta})\\
&\approx&[1-4\epsilon^2c_{\lam}^2\sin^2\frac{\zeta}{2}]
e^{i2c_{\lam}^2\epsilon^2(\zeta-\sin\zeta)}\\
X&=&i\epsilon s_{\lam}(1-e^{-i\zeta})\\
Y&=&i\epsilon c_{\lam}(1-e^{-i\zeta})
\end{array}
\end{equation}
an initial laser beam of $\Psi(0)=(c_{\theta},s_{\theta},0)^T$
will evolve into the state \begin{equation} \Psi_{N,{\rm
round}}=e^{i2\omega NL}\left(
\begin{array}{c}
\left[1-4N\epsilon^2s_{\lam}^2\sin^2\frac{\zeta}{2}\right]
e^{i2N\epsilon^2s_{\lam}^2(\zeta-\sin\zeta)}\\
\left[1-4N\epsilon^2c_{\lam}^2\sin^2\frac{\zeta}{2}\right]
e^{i2N\epsilon^2c_{\lam}^2(\zeta-\sin\zeta)}
\end{array}\right)
\end{equation}
for $N\epsilon^2\ll 1$. The induced rotation and ellipticity are
\begin{eqnarray}
\Delta\theta&=&-2N\epsilon^2\sin^2\frac{\zeta}{2}\sin 2\theta
\cos 2\theta_{\lam}\\
\chi&=&N\epsilon^2(\zeta-\sin\zeta)\sin 2\theta\cos 2\theta_{\lam}
\label{round}
\end{eqnarray}
which are modified by a factor $\cos 2\theta_{\lam}$ from the
standard results for a particle of definite parity. This simple
modification means that the potential tension between the PVLAS
result on rotation (pointing to a pseudoscalar) \cite{PVLAS05} and
its preliminary result on ellipticity (favoring a scalar)
\cite{PVLAS06} is not relaxed in the parity non-conserving case.
When the two interactions are of the same strength, i.e., $\cos
2\theta_{\lam}=0$, no net optical effects remain as intuitively
expected, although the photon to $\varphi$ transition generally
occurs. It is also clear that a reflection symmetric experiment
like PVLAS cannot tell a pseudoscalar or scalar particle from one
of no definite parity.

If our aim is to work out the rotation and ellipticity at the
first non-trivial order in $\epsilon$, there is a simpler way to
proceed. The amount collected on the return trip can be obtained
from that on the forward trip by $c_{\lam}\to -c_{\lam}$, which is
equivalent to $\theta_{\lam}\to -\theta_{\lam}$ in eqns.
(\ref{rotation},\ref{ellipticity}). Adding the amount for a round
trip yields the result shown in eqn (\ref{round}) for $N=1$.

To summarize, we have studied the general electromagnetic
couplings of a neutral, spin-zero particle in an effective field
theory at the energy scale of order eV. Our consideration was
based on the observation that, if the PVLAS results are explained
in terms of an axion-like particle, the latter cannot be a
QCD-like axion. In this circumstance, there is no theoretically
strong preference for a pseudoscalar or scalar particle; instead,
it could be a particle of no definite parity. We have considered
the effects of such a particle on the evolution of a light beam
propagating in a transverse magnetic field, and calculated the
changes in the light polarization for two experimental set-ups.
For a parity asymmetric set-up, e.g., with light propagating in
one direction, interesting phenomena can occur for special
polarization configurations where neither rotation nor ellipticity
is induced although particle transitions still take place. Those
configurations are determined by the relative strength of the
interactions. In a parity symmetric set-up like PVLAS, however,
optical changes are simply modified by a common factor, compared
to the pure-parity case. The factor is again fixed by the relative
strength of the interactions. Thus, if the results in such an
experiment can be explained in terms of a particle with definite
parity, they can be equally well explained in terms of a particle
of no definite parity by adjusting the factor. Nevertheless, the
parity property could be studied in a photon regeneration
experiment if the polarization of regenerated photons could be
measured.

In optical experiments like BFRT and PVLAS, a very small amount of
gas is usually introduced into the cavity for calibration
purposes. This amounts to introducing new terms in our $\Omega$
shown in eqn.(\ref{linear}) to replace the diagonal zeros (due to
Cotton-Mouton effect) and off-diagonal ones (for Faraday effect in
the presence of a magnetic field component along the propagation
direction). This more general mixing case is also relevant for
phenomena in the stellar environment. The interplay with the
simple mixing scheme discussed in this work, including possible
modifications in pressure, magnetic field and path length
dependencies, deserves further study to which we hope to come back
soon.

{\bf Acknowledgements} I would like to thank the anonymous referee
for many useful suggestions and comments that have helped me
clarify some points in the original version of the paper.

%\newpage
%\baselineskip=20pt


\begin{thebibliography}{30}
%\begin{enumerate}

\bibitem{heisenberg}W. Heisenberg, H. Euler, Z. Phys. 98 (1936)
714; V.F. Weisskopf, Mat. Fys. Medd.-K. Dan Vidensk. Selsk. 14
(1936) 6; J. Schwinger, Phys. Rev. 82 (1951) 664

\bibitem{adler70}S.L. Adler, J.N. Bahcall, C.G. Callan, M.N.
Rosenbluth, \PRL 25 (1970) 1061; S.L. Adler, Ann. Phys. (N.Y.) 67
(1971) 599

\bibitem{iacopini79}E. Iacopini, E. Zavattini, \PL 85B (1979) 151
% Experimental method to detect the vacuum birefringence induced by
% a magnetic field

\bibitem{BFRT}BFRT Collaboration, R. Cameron {\it et al.}, \PR D47
(1993) 3707
% Search for nearly massless, weakly coupled particles by optical techniques

\bibitem{PVLAS05}PVLAS Collaboration, E. Zavattini {\it et al.},
\PRL 96 (2006) 110406 [hep-ex/0507107]

\bibitem{PVLAS06}G. Cantatore for PVLAS Collaboration, talk at IDM
2006, Rhodos, Greece, Sept. 11-16th, 2006.

\bibitem{adler06}S.L. Adler, J. Phys. A40 (2007) F143 [hep-ph/0611267]

\bibitem{biswas06}S. Biswas, K. Melnikov, \PR D75 (2007) 053003
[hep-ph/0611345]

\bibitem{hu07}X.-P. Hu, Y. Liao, hep-ph/0702111

\bibitem{sikivie83}P. Sikivie, \PRL 51 (1983) 1415; \PRL 52 (1984)
695 (Erratum)
% suggesting the helioscape idea used by CAST

\bibitem{maiani86}L. Maiani, R. Petronzio, E. Zavattini, \PL B175
(1986) 359

\bibitem{raffelt88}G. Raffelt, L. Stodolsky, \PR D37 (1988) 1237

\bibitem{cast05}CAST Collaboration, K. Zioutas {\it et al.}, \PRL 94
(2005) 121301 [hep-ex/0411033]

\bibitem{cast07}CAST Collaboration, S. Andriamonje {\it et al.},
hep-ex/0702006

\bibitem{raffelt06}For a review, see e.g.: G. Raffelt, hep-ph/0611118

\bibitem{conflict}An incomplete list includes, %
P. Brax, C. van de Bruck, A.-Ch. Davis, hep-ph/0703243; %
%Has PVLAS detected the chameleon?
R. Foot, A. Kobakhidze, hep-ph/0702125; %
%A Simple explanation of the PVLAS anomaly in spontaneously broken mirror models.
P. Jain, S. Stokes, hep-ph/0611006;
% Self Interacting Dark Matter in the Solar System
% strongly interacting scalar pervades in galaxy and weakly interacts with
% matter to evade astrophysical constraints
J. Jaeckel, E. Masso, J. Redondo, A. Ringwald, F. Takahashi, \PR
D75 (2007) 013004 [hep-ph/0610203];
% The Need for purely laboratory-based axion-like particle searches
% stellar suppresion of axion production due to environmental effects
R.N. Mohapatra, S. Nasri, \PRL 98 (2007) 050402 [hep-ph/0610068]; %
%Reconciling the CAST and PVLAS results.
P. Jain, S. Mandal, Int.J.Mod.Phys.D15 (2006) 2095
[astro-ph/0512155];
% Evading the astrophysical limits on light pseudoscalars.
% Scalars trapped in the sun due to strong self-interaction
E. Masso, J. Redondo, JCAP 0509 (2005) 015 [hep-ph/0504202]
% Evading astrophysical constraints on axion-like particles.
% axion-like particles trapped in the sun; suppressed Primakoff
% process in stellar medium

\bibitem{milli}H. Gies, J. Jaeckel, A. Ringwald, \PRL 97 (2006)
140402 [hep-ph/0607118];
% Polarized light propagating in a magnetic field as a probe for
% millicharged fermions
M. Ahlers, H. Gies, J. Jaeckel, A. Ringwald, \PR D75 (2007) 035011
[hep-ph/0612098]
% On the particle interpretation of the PVLAS data: neutral versus charged
% particles

\bibitem{holdom86}B. Holdom, \PL B166 (1986) 196

\bibitem{abel06}S.A. Abel, J. Jaeckel, V.V. Khoze, A. Ringwald, hep-ph/0608248
%Illuminating the Hidden Sector of String Theory by Shining Light through
%a Magnetic Field.

\bibitem{masso06}E. Masso, J. Redondo, \PRL 97 (2006) 151802
[hep-ph/0606163]
%Compatibility of CAST search with axion-like interpretation of PVLAS results.

\bibitem{gies06}H. Gies, J. Jaeckel, A. Ringwald,
Europhys. Lett. 76 (2006) 794 [hep-ph/0608238]
%Accelerator Cavities as a Probe of Millicharged Particles.

\bibitem{Gninenko06}S.N. Gninenko, N.V. Krasnikov, A. Rubbia, hep-ph/0612203
%Search for millicharged particles in reactor neutrino experiments:
%A Probe of the PVLAS anomaly.

\bibitem{regen_e}K. Van Bibber, {\it et al.}, \PRL 59 (1987) 759;
% Proposed experiment to produce and detect light pseudoscalars
G. Ruoso, {\it et al.}, Z. Phys. C56 (1992) 505
% Search for photon regeneration in a magnetic field

\bibitem{regen_t}For recent discussions on photon regeneration, see e.g.: R.
Rabadan, A. Ringwald, K. Sigurdson \PRL 96 (2006) 110407
[hep-ph/0511103];
% Photon regeneration from pseudoscalars at X-ray laser facilities
M. Fairbairn, T. Rashba, S. Troitsky, astro-ph/0610844;
% Shining light through the Sun
P. Sikivie, D.B. Tanner, K. van Bibber, hep-ph/0701198
% Resonantly enhanced axion-photon regeneration

\bibitem{alps}K. Ehret {\it et al.}, hep-ex/0702023

\bibitem{hill88}C.T. Hill, G.G. Ross, \NP B311 (1988) 253

%\end{enumerate}
\end{thebibliography}
\end{document}